\documentclass{article}

\usepackage{PRIMEarxiv}

\usepackage[utf8]{inputenc} % allow utf-8 input
\usepackage[T1]{fontenc}    % use 8-bit T1 fonts
\usepackage{url}            % simple URL typesetting
\usepackage{booktabs}       % professional-quality tables
\usepackage{amsfonts}       % blackboard math symbols
\usepackage{nicefrac}       % compact symbols for 1/2, etc.
\usepackage{microtype}      % microtypography
\usepackage{lipsum}
\usepackage{fancyhdr}       % header

\usepackage{multirow} % for tables
\usepackage{booktabs} % for table formatting
\usepackage{wrapfig} % floating figure embedded in text
\usepackage{bm} % using math bold equations
\usepackage{amsmath} % for math symbols
\usepackage{float}
\usepackage{tabularx}
\usepackage{algorithm}
\usepackage{algpseudocode}

\usepackage[pdftex]{graphicx}
\usepackage[colorlinks=true,linkcolor=blue,citecolor=blue,urlcolor=blue]{hyperref}
\graphicspath{{media/}}     % organize your images and other figures under media/ folder

\title{Modeling Autonomous Shifts Between Focus State and Mind-Wandering Using a Predictive-Coding-Inspired Variational RNN Model
}

\author{
  Henrique Oyama\\
  Cognitive Neurorobotics Research Unit \\
  Okinawa Institute of Science and Technology Graduate University \\
  Okinawa\\
  \texttt{henrique.oyama@oist.jp} \\
   \And
  Jun Tani \\
  Cognitive Neurorobotics Research Unit \\
  Okinawa Institute of Science and Technology Graduate University \\
  Okinawa\\
  \texttt{jun.tani@oist.jp} \\
}

\begin{document}
\maketitle

%===================================================
%===================================================
\begin{abstract}
%\lipsum[1]
The current study investigates possible neural mechanisms underling autonomous shifts between focus state and mind-wandering by conducting model simulation experiments.
On this purpose, we modeled perception processes of continuous sensory sequences using our previous proposed variational RNN model which was developed based on the free energy principle.
The current study extended this model by introducing an adaptation mechanism of a meta-level parameter, referred to as the meta-prior $\mathbf{w}$, which regulates the complexity term in the free energy.
Our simulation experiments demonstrated that autonomous shifts between focused perception and mind-wandering take place when $\mathbf{w}$ switches between low and high values associated with decrease and increase of the average reconstruction error over the past window. In particular, high $\mathbf{w}$ prioritized top-down predictions while low $\mathbf{w}$ emphasized bottom-up sensations. This paper explores how our experiment results align with existing studies and highlights their potential for future research.
\end{abstract}

% keywords can be removed

%===================================================
%===================================================
%\input{sections/1_Introduction} 
\section{Introduction}\label{sec:1:introduction}

During mindfulness practice, such as focusing on sensations like breathing, our attention sometimes spontaneously deviates to mental imagery or thoughts about the past and future, a phenomenon known as mind-wandering \cite{smallwood2015science,christoff2016mind,seli2016mind}. 
This shift from a focused state to mind-wandering can occur not only during meditation but also in everyday activities, such as driving, listening to music, or tasting food. 

Mind-wandering tends to occur more frequently during tasks that are either too easy or too difficult. 
When tasks are less demanding, such as simply attending to breathing, instances of mind-wandering increase. 
Conversely, during more challenging tasks, like reading complex material, our minds are more prone to wander because maintaining focus becomes difficult over extended periods \cite{seli2018role,peral2023intentional}. 

An interesting aspect is that the transition from the focused state (FS) to the mind-wandering state (MW) often happens without conscious awareness, whereas the shift from MW back to FS involves recognizing the mind-wandering episode consciously \cite{smallwood2013not}. 
Various studies have investigated the psychological and systematic mechanisms underlying these shifts. 
For example, \cite{henriquez2016fluctuating} argued that the transition from FS to MW is gradual, as evidenced by increasing response times during focused tasks. 
In contrast, \cite{vago2016brain} suggested that the shift is abrupt, triggered by sudden internal or external stimuli.

\cite{voss2018new} proposed a model where mental states alternate between FS and MW, with MW episodes ending when individuals consciously recognize their mind-wandering and return to the task. 
This ``two-stage model'' assumes that the probability of being in FS is higher at the beginning of an episode and decreases over time. 
However, contrary to this prediction, \cite{zukosky2021spontaneous} found that the probability of FS does not decline within an FS-MW episode in a subject study introducing a probe in a random timing during the episode. 
To address this discrepancy, the authors proposed the ``multiple sub-event model'', which hypothesizes that unconscious alternations between FS and MW occur multiple times before an individual becomes aware of being in MW. 
Their simulation study suggested that as the number of sub-sequences increases, the decline in the probability of FS becomes less pronounced.
%While these studies have elucidated some aspects of the FS to MW shift through psychological observation, they have not adequately explained the underlying neuronal mechanisms.  

Although the above mentioned studies clarified some phenomena in the shift from FS to MW from psychological observation, they have not provided sufficient accounts for the underlying neuronal mechanisms.
%% Some studies about resting state and MW
Recently, some studies \cite{sandved2021towards,idei2024awareness} suggested system level neuroscience models incorporated with the concept of the free energy principle (FEP) \cite{friston2005theory}.
Here, FEP is briefly explained for better understanding of the readers.
The FEP is a neuroscience theory that has attracted large attention. 
The FEP posits that humans and animals execute various functions such as learning, perception, and action generation to maximize their chances of survival by minimizing surprises they encounter during interaction with the environment. 
According to the FEP, these functions are achieved by optimizing generative models for predicting the sensation, whereby a common statistical quantity called free energy is minimized. 
The FEP supports two frameworks, one is predictive coding and the other is active inference.
Predictive coding provides a formalism accounting for how agents perceive sensations. 
It suggests that the brain predicts sensory observations in the top-down pathway, while at the same time updating posterior beliefs about those sensations in the bottom-up pathway whenever errors arise between predictions and observations \cite{rao1999, friston2005theory, clark2015}. 
By updating posterior beliefs in the direction of minimizing errors, perceptual inference for the observed sensation can be achieved. 
On the other hand, active inference (AIF) provides a theory for action generation by assuming that the brain is embodied deeply and embedded in the environment, such that acting on it changes future sensory observation. 
Then, AIF considers that actions should be selected such that the error between the desired and predicted sensations can be minimized \cite{friston2010action, friston2011}.

\cite{sandved2021towards} postures the underlying mechanism of shift from FS to MW using active inference of ``mental action'' in terms of attention changes.
The proposed model assumes hierarchical probabilistic generative model wherein the hidden meta-awareness states in the higher level account for "how aware am I of where my attentions is?", the hidden mental states in the middle level dealing with focus of attention account for "what am I paying attention to?", and the sensorimotor hidden states in the lowest level do for "what am I perceiving or trying to do?" according to the authors.
The states at each level condition the ones in the next lower level by controlling their precisions or beliefs.
Agent's perceptual and attentional states are inferred at each time step by means of active inference in minimizing the expected free energy.
The results of simulation experiments show that when the meta-awareness state is manually shifted from high to low, distracted or MW state is developed more frequently.
Under this condition, redirection back to FS by consciously being aware of the current MW state tends to take more time because of less precision in the attention toward distracted state.

\cite{idei2024awareness} investigated mind-wandering mechanism by conducting a model simulation study on allostasis using hierarchically organized variation recurrent neural network, so-called the PV-RNN \cite{Reza2019}. 
Dynamic behavior of PV-RNN can be characterized by a meta-level parameter, referred to as meta-prior $\mathbf{w}$, that regulates the complexity term against the accuracy term in free energy which is minimized in the inference of the posterior probability distribution of the latent variables.
It was shown that high setting of meta-prior $\mathbf{w}$ enhances generation of the top-down imagery while low setting of it enhances the bottom-up sensory perception \cite{wataru2020,chameFrontier2020,wirkuttis2023turn}.
Analogous to this, \cite{idei2024awareness} showed that low setting of $\mathbf{w}$ generates stronger sensory bottom-up which leads to FS wherein less change in movement as well as neural activity are observed. 
On the other hand, high setting of $\mathbf{w}$ generates weak attention to sensation and stronger top-down which leads to MW wherein more movement as well as neural activity.

The aforementioned FEP-based studies provide valuable insights into macroscopic neural mechanisms, such as redirecting attention to focused states by inferring one’s attentional state, or generating mind-wandering by balancing top-down and bottom-up information flows. However, these studies do not provide systematic explanations for how the shifts between FS and MW could be autonomously generated, since the shift from FS to MW in \cite{sandved2021towards} is caused by manual change of the meta-aware state from high to low and the one in \cite{idei2024awareness} does this by resetting meta-prior $\mathbf{w}$ from low to high value. 

In this regard, the current study speculates that autonomous transition between FS and MW could be generated by introducing an adaptation mechanism of meta-prior $\mathbf{w}$ to PV-RNN in which $\mathbf{w}$ is modulated with response to some macroscopic variables such as an average prediction error. 
In our study, PV-RNN learns to predict a target sequence of continuously changing sensory patterns which is generated by means of predetermined probabilistic transitions among a set of cyclic patterns.
In the test phase after the training, given one of the pre-trained cyclic patterns as the target inputs, the PV-RNN predicts encountering sensory inputs by simultaneously inferring the approximated posterior of the latent state at each time step by minimizing the free energy while adapting $\mathbf{w}$.
Analogous to studies \cite{wataru2020,chameFrontier2020,wirkuttis2023turn,idei2024awareness}, when $\mathbf{w}$ modulates to a lower value by reflecting surge of the average reconstruction error, the inference process may improve by placing greater emphasis on bottom-up sensations. This situation may correspond to FS.
On the other hand, when $\mathbf{w}$ modulates to a higher value by responding to decline of the average reconstruction error, the PV-RNN may generate top-down imagery by following the learned probabilistic transitions of patterns while ignoring the target sensory inputs.
This may correspond to MW.
Our simulation study with PV-RNN under various parameter settings will evaluate this hypothesis.
The following section introduces the proposed model, followed by a detailed description of the simulation experiment setup, the presentation of the results, and a discussion that includes proposals for extensions to future research work.

%===================================================
%===================================================
%\input{sections/2_Methods} 
\section{Materials and Methods}\label{sec:2}
%%%
\subsection{Overview}\label{sec:2:overview}
%%%
This study investigates autonomous shifts between the focused state (FS) and mind-wandering (MW) during a perception task using sequential sensory input patterns. 
The predictive coding framework is employed to model this perception process. 
Predictive coding assumes a generative model that predicts sensory sequences by learning both the latent state transition function and the likelihood mapping from latent states to sensory observations. 
Additionally, this generative model infers the current latent state through continuous sensory sequence observations. 

Both learning and inference processes are achieved by minimizing prediction error or, more specifically, free energy. 
We hypothesize that FS is enhanced by strengthening bottom-up inference, while MW becomes more likely by emphasizing top-down sensory pattern generation. 
It is also hypothesized that shifts between FS and MW take place autonomously incorporating with adaptation of meta-level states with response to particular system variables.
To test this, we propose an extended version of a variational recurrent neural network model, referred to as the Predictive Coding Inspired Variational RNN (PV-RNN) \cite{Reza2019}. 
Details of the original PV-RNN and its extensions are provided in the following sections.

%%%
\subsection{Predictive Coding Inspired Variational RNN Model (PV-RNN)}\label{sec:2:rnn}
%%%
The PV-RNN is based on the free energy principle \cite{friston2005theory}, where learning and inference are achieved by minimizing free energy (Equation \ref{eq:F}) in accordance with Bayes' theorem:

\begin{equation}
\begin{aligned}\label{eq:F}
    \mathcal{F} = \underbrace{D_{\rm KL}[q_\phi(\mathbf{z}|\mathbf{X})\Vert p_\theta(\mathbf{z})]}_{\text{complexity}} - \underbrace{\mathbb{E}_{q_\phi(\mathbf{z}|\mathbf{X})}[\log p_\theta(\mathbf{X}|\mathbf{z})]}_{\text{accuracy}}
\end{aligned}
\end{equation}

Here, \( p_\theta(\mathbf{X}) \) is the marginal likelihood of the sensory observation \( \mathbf{X} \), given the generative model \( p_\theta \) parameterized by \( \theta \). 
The latent variables \( \mathbf{z} \) and inference model \( q_\phi \), parameterized by \( \phi \), allow for posterior inference through minimization of free energy. 
Free energy consists of two terms: the complexity term (a measure of divergence between prior and posterior distributions) and the accuracy term (log-likelihood of sensory observations) \cite{Fri10b}.
PV-RNN serves as both a generative model and an inference model. 
The generative model predicts future sensory inputs via top-down processes, while the inference model estimates the approximate posterior from observed sensory sequences through free energy minimization as bottom-up processes. 

The following subsections describe the PV-RNN implementation and the use of the meta-prior \( \mathbf{w} \).

%%%
\subsubsection{Model Implementation}\label{sec:2:architecture}
%%%
The free energy \(\tilde{\mathcal{F}}\) for PV-RNN predicting a time series of \( T \) steps is given by:

\begin{equation}
\begin{split}
    {\mathcal{F}} =& \underbrace{\mathbf{w} 
    \sum_{t=1}^T \mathbb{E}_{q_\phi(\mathbf{z}_{1:t-1}|\mathbf{d}_{t-1},\mathbf{X}_{t-1:T})} \big[D_{\text{KL}}[q_\phi(\mathbf{z}_t|\mathbf{d}_{t-1},\mathbf{X}_{t:T})\Vert p_\theta(\mathbf{z}_t|\mathbf{d}_{t-1})]\big]}_{\text{complexity}}\\
    & - \underbrace{\sum_{t=1}^T\mathbb{E}_{q_\phi(\mathbf{z}_{1:t-1}|\mathbf{d}_{t-1},{\mathbf{X}}_{t:T})}[\log p_\theta(\mathbf{X}_t|\mathbf{d}_t)]}_{\text{accuracy}}
\end{split}
\label{eq:freeenergy}
\end{equation}

\begin{figure}[htbp]
\centering
\includegraphics[width=0.9\textwidth]{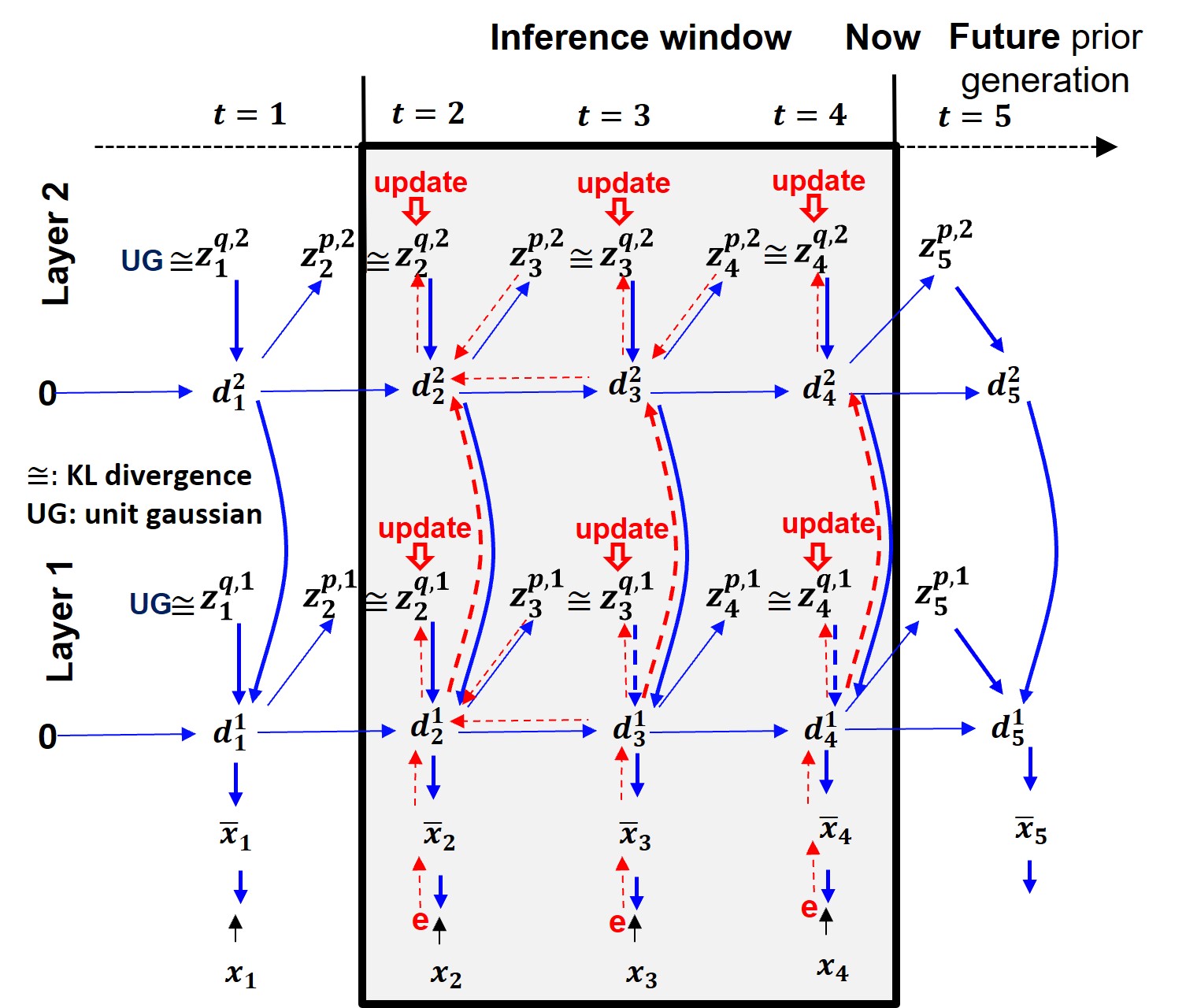}
\caption{A hierarchical two-layer PV-RNN architecture. Solid blue lines represent the generative process, while dotted red lines indicate the inference process. The shaded area shows an inference window of length 3. 
% The meta-prior \( \mathbf{w} \) weights the complexity term in the inference process. 
\label{fig:PV-RNN}
}
\end{figure}

PV-RNN introduces two types of latent variables: probabilistic latent variables (\(\mathbf{z}\)) governed by Gaussian distributions, and deterministic latent variables (\(\mathbf{d}\)). Their relationships are shown in Figure \ref{fig:PV-RNN}.
In equation \ref{eq:freeenergy}, a meta-level parameter, named meta-prior \( \mathbf{w} \), is introduced to balance the complexity and accuracy terms during this process. 
This regulation is particularly important when the limited amount of training data prevents reliable estimation of latent variable distributions. 
Also, dynamic behavior of PV-RNN is largely affected by setting of the meta-prior.
It was shown that high setting of meta-prior $\mathbf{w}$ enhances generation of the top-down imagery while low setting of it enhances the bottom-up sensory perception \cite{wataru2020,chameFrontier2020,wirkuttis2023turn,idei2024awareness}.

Next, the forward computation of each variable used in PVRNN is described.
At each time step \( t \), the internal states of the \( l \)-th layer (\( \mathbf{h}_t^l \)) are recursively computed:

\begin{equation}\label{eq:internalstates}
\begin{aligned}
    \mathbf{h}^l_t &= \left(1 - \frac{1}{\tau^l}\right)\mathbf{h}^l_{t-1} +
    \frac{1}{\tau^l} \left( \mathbf{W}^{ll}_{dd}\mathbf{\tilde{d}}^l_{t-1} + \mathbf{W}^{ll}_{zd}\mathbf{z}^l_t + \mathbf{W}^{ll+1}_{dd}\mathbf{\tilde{d}}^{l+1}_{t-1} + \mathbf{W}^{ll-1}_{dd}\mathbf{\tilde{d}}^{l-1}_{t-1} + \mathbf{b}_h^l \right)\\
    \mathbf{\tilde{d}}^l_t &= \tanh(\mathbf{h}^l_t)
\end{aligned}
\end{equation}

The PV-RNN structure supports hierarchical information processing using time constants \( \tau^l \), enabling the differentiation of temporal dynamics across layers \cite{yamashita2008, Schillaci2020}. 

The generative model computes prior distributions (\( \mathbf{z}^p_t \)) as Gaussian variables parameterized by mean (\( \bm{\mu}_t^p \)) and standard deviation (\( \bm{\sigma}_t^p \)):
\begin{equation}\label{eq:p}
\begin{aligned}
    \bm{\mu}^p_t &= \tanh(\mathbf{W}^{ll}_{d\mu}\mathbf{\tilde{d}}_{t-1} + \mathbf{b}_\mu^p)\\
    \bm{\sigma}^p_t &= \exp(\mathbf{W}^{ll}_{d\sigma}\mathbf{\tilde{d}}_{t-1} + \mathbf{b}_\sigma^p)\\
    \mathbf{z}^p_t &= \bm{\mu}^p_t + \bm{\sigma}^p_t * \bm{\epsilon}_t \quad \text{with } \bm{\epsilon}_t \sim \mathcal{N}(\mathbf{0}, \mathbf{I})
\end{aligned}
\end{equation}
$\mathbf{b}^p_\mu$ and $\mathbf{b}^p_\sigma$ are bias terms for $\bm{\mu}^p_t$ and $\bm{\sigma}^p_t$, respectively. 
$\bm{\epsilon}$ represents a noise sampled from a standard normal distribution for usage of the reparameterization trick \cite{kingma2014autoencoding}. 
Analogous to the computation of the prior distribution, the inference model $q_\phi$ approximates the posterior $\bm{z}^q_t$ as a Gaussian distribution with mean $\bm{\mu}_t^q$ and standard deviation $\bm{\sigma}_t^q$.

% Eq.5
\begin{equation} 
\begin{aligned}
\label{eq:posterior}
    \bm{\mu}^q_t &= \tanh(\mathbf{W}^{ll}_{d\mu}\mathbf{\tilde{d}}_{t-1} + \mathbf{A}^\mu_t + \mathbf{b}_\mu^q)\\
    \bm{\sigma}^q_t &= \exp(\mathbf{W}^{ll}_{d\sigma}\mathbf{\tilde{d}}_{t-1} + \mathbf{A}^\sigma_t + \mathbf{b}_\sigma^q) \\
    \mathbf{z}^q_t &= \bm{\mu}^q_t + \bm{\sigma}^q_t * \bm{\epsilon}_t \quad \textrm{with }\bm{\epsilon}_t\sim\mathcal{N}(\mathbf{0},\mathbf{I})
\end{aligned}
\end{equation}
where $\mathbf{b}^q_\mu$ and $\mathbf{b}^q_\sigma$ are bias terms for computing $\bm{\mu}^q_t$ and $\bm{\sigma}^q_t$, respectively. $\mathbf{A}^\mu_t$ and $\mathbf{A}^\sigma_t$ are adaptive variables to be optimized for inferring the posterior distribution which is parameterized by $\bm{\mu^q_t}$ and $\bm{\sigma}^q_t$.

Intuitively, the random variable $\mathbf{z}^p$ can be regarded as a time-dependent prior/top-down expectation about the encountering sensation. 
The adaptive vector $\mathbf{A}$ (i.e., $\mathbf{z}^q$) can be regarded as the approximate posterior distribution that may or may not be close to the prior distribution, depending on the setting of meta-prior. 
$\mathbf{z}^p$ and $\mathbf{z}^q$ are used by the generative and inference model, respectively to compute the latent variable $\mathbf{d}$. 

%%%
\subsubsection{Learning and Inference} \label{sec:2:loss}
%%%
The free energy $\mathcal{F}$ of PV-RNN can be computed as follows by adapting the original equation \ref{eq:freeenergy}. 
Given a PV-RNN with $L$ layers, predicting a $T$ time series sensory inputs, $\mathcal{F}$ can be written as
% Eq.6
\begin{equation}
    \begin{split}
        \mathcal{F}=&\sum_{t=1}^T\Bigg[\sum_{l=1}^L\tilde{\mathbf{w}}^lD_\text{KL}[q_\phi(\mathbf{z}^l_t|\mathbf{d}^l_{t-1},\mathbf{X}_{t:T})\Vert p_\theta(\mathbf{z}^l_t|\mathbf{d}^l_{t-1})]\Bigg] -\sum_{t=1}^T\Vert\mathbf{X}_t-\bar{\mathbf{X}}_t\Vert^2_2
    \end{split}
    \label{eq:f1}
\end{equation}
where $\tilde{\mathbf{w}}^l$ is $\mathbf{w}$ specific to $l$th layer, and $\bar{\mathbf{X}}$ denotes the prediction output of the PV-RNN. 
In equation \ref{eq:f1}, we approximate the expectation with respect to the approximate posterior by iterative sampling. 
Also, the accuracy term is replaced by the squared error, which can be regarded a special case of computation of log-likelihood wherein each dimension of $\mathbf{X}$ and $\mathbf{\bar{X}}$ is independent and follows a Gaussian distribution with standard deviation $1$. 
Since the Kullback-Leibler (KL) divergence between two one-dimensional Gaussian distributions takes a simple expression, equation \ref{eq:f1} is reduced to
% Eq.7
\begin{equation}
\mathcal{F}=\sum_{t=1}^T\Bigg[\sum_{l=1}^L\tilde{\mathbf{w}}^l\sum_{r=1}^{R^l_z}\delta(l,r,t)\Bigg]-\sum_{t=1}^T\Vert\mathbf{X}_t-\mathbf{\bar{X}}_t\Vert^2_2
\end{equation}
where
% Eq. 8
\begin{equation}
    \delta(l,r,t)=\log\frac{\sigma^{p,l,r}_t}{\sigma^{q,l,r}_t}+\frac{(\mu^{q,l,r}_t-\mu^{p,l,r}_t)^2+(\sigma^{q,l,r}_t)^2}{2(\sigma^{p,l,r}_t)^2}-\frac{1}{2}
\end{equation}
%
% Henrique should make sure that the following normalization is conducted in the current program.
$\mu^{p,l,r}_t$ represents $r$th element of $\bm{\mu}^l_t$ of the prior, and the same notation is applied to $\mu^{q,l,r}_t$, $\sigma^{p,l,r}_t$, and $\sigma^{q,l,r}_t$. $R^l_z$ denotes the dimension of $\mathbf{z}^l_t$. 
Given that the complexity term is summed over all the dimension of $\mathbf{z}$, which is arbitrary to the network design, and the accuracy term is to all the data dimension, which varies among data, the free energy is normalized with respect to the dimension of $\mathbf{z}$ and the data dimension. 
Therefore, introducing such normalization, the free energy of PV-RNN in the study is computed by
\begin{equation} 
\begin{aligned}%\footnotesize
    \mathcal{F} 
    &= \underbrace{\sum_{t=1}^T \Bigg[
     \sum_{l}^L \frac{\mathbf{w}^l}{R_z^l} \delta(l,r,t)\Bigg]}_{\rm complexity} - \underbrace{\frac{1}{R_X} \Bigg[\sum_{t=1}^T  \|\mathbf{X}_t - {\bar{\mathbf{X}}_t \|^2_2}\Bigg]}_{\rm accuracy}
\label{eq:loss}
\end{aligned}
\end{equation}
where $R_X$ is the data dimension, $R^l_z$ is the number of $\mathbf{z}$ variables in each layer, and $\mathbf{w}^l=R_z^l\tilde{\mathbf{w}}^l$.

By minimizing equation \ref{eq:loss}, the posterior inference is performed during network learning and during the perception task. 
Figure \ref{fig:PV-RNN} shows a schematic illustration of the posterior inference process of a two-layer PV-RNN model used in the current simulation with an optimization window of three time steps. 
At every sensory step, an adaptive variable $\mathbf{A}$ in the window is optimized through multiple epochs of stochastic gradient descent.
In the network learning phase, weights and bias parameters $\theta$ and $\phi$ of the generative and inference models, including an adaptive variable $\mathbf{A}$ for the approximate posterior $\bm{z}^q$ are jointly optimized. 
In the perception task phase, network parameters $\theta$ and $\phi$ are fixed, and free energy is minimized at each time step within a dedicated inference window by optimizing only $\mathbf{A}$ parameterizing the approximate posterior.

%%%
\subsubsection{Adaptation of Meta-Prior} \label{sec:2:w-adaptation}
%%%
The meta-prior \( \mathbf{w} \) is dynamically adapted based on the average prediction error (\( \mathbf{er_{sum}} \)) over a fixed length time window in the past. 
When the error decreases below a predefined threshold (\( \mathbf{Thr_{L}} \)), \( \mathbf{w} \) transitions to a high value ($\mathbf{w}^{H}$), prioritizing top-down generation, which leads to generating MW.
This can be intuitively understood from analogy that continuing easy or predictable tasks tends to initiate MW \cite{peral2023intentional,seli2018role}.
Conversely, when the average prediction error exceeds an upper threshold (\( \mathbf{Thr_{H}} \)), \( \mathbf{w} \) transitions to a low value ($\mathbf{w}^{L}$), enhancing bottom-up inference. 
The implementation strategy for autonomous meta-prior switching between FS and MW is described in Algorithm \ref{alg:adaptive_switching}. Specifically, the probabilistic shifting between the two modes is given by equations \ref{eq:adaptW_F}-\ref{eq:adaptW_MW}, where $Temp$ is the temperature, a tunable parameter that can reflect how stochastic or deterministic the system is (see Section \ref{sec:results:testing}). 
It is highly speculated that this dynamic adaptation should enable autonomous transitions between FS and MW, as will be validated in the simulation experiments detailed in subsequent sections.

\begin{algorithm}[H]
\caption{Autonomous Meta-Prior Switching Between Focus State (FS) and Mind-Wandering (MW)}
\label{alg:adaptive_switching}
\begin{algorithmic}[1]
%\State \textbf{Input:} Cumulative error sum $er_{sum}$, thresholds $Thr_L$, $Thr_H$, temperature $T$
\State Initialize meta-prior $\mathbf{w}$ (either $\mathbf{w}^{L}$ or $\mathbf{w}^{H}$)

\If {$\mathbf{w}$ == $\mathbf{w}^{L}$} 
    \State Compute transition probability from FS to MW:
    \begin{equation}
        \mathcal{P}(FS \rightarrow MW) = \text{sigmoid}\left(\frac{-(er_{sum} - Thr_L)}{Temp}\right)
        \label{eq:adaptW_F}
    \end{equation}
    \State Generate random number $r \sim \mathcal{G}(0, 1)$
    \If {$r < \mathcal{P}(FS \rightarrow MW)$}
        \State Set meta-prior to $\mathbf{w} \leftarrow \mathbf{w}^{H}$
    \EndIf
\ElsIf {$\mathbf{w}$ == $\mathbf{w}^{H}$} 
    \State Compute transition probability from MW to FS:
    \begin{equation}
        \mathcal{P}(MW \rightarrow FS) = \text{sigmoid}\left(\frac{er_{sum} - Thr_H}{Temp}\right)
        \label{eq:adaptW_MW}
    \end{equation}
    \State Generate random number $r \sim \mathcal{G}(0, 1)$
    \If {$r < \mathcal{P}(MW \rightarrow FS)$}
        \State Set meta-prior to $\mathbf{w} \leftarrow \mathbf{w}^{L}$
    \EndIf
\EndIf

%\State \textbf{Output:} Updated meta-prior state $w$
\end{algorithmic}
\end{algorithm}

%===================================================
%===================================================
%\input{sections/3_Results}
\section{Experiments and Results}\label{sec:results}
\subsection{Model Training}\label{sec:results:training}
\begin{figure}[H]
\centering
\includegraphics[width=0.9\textwidth]{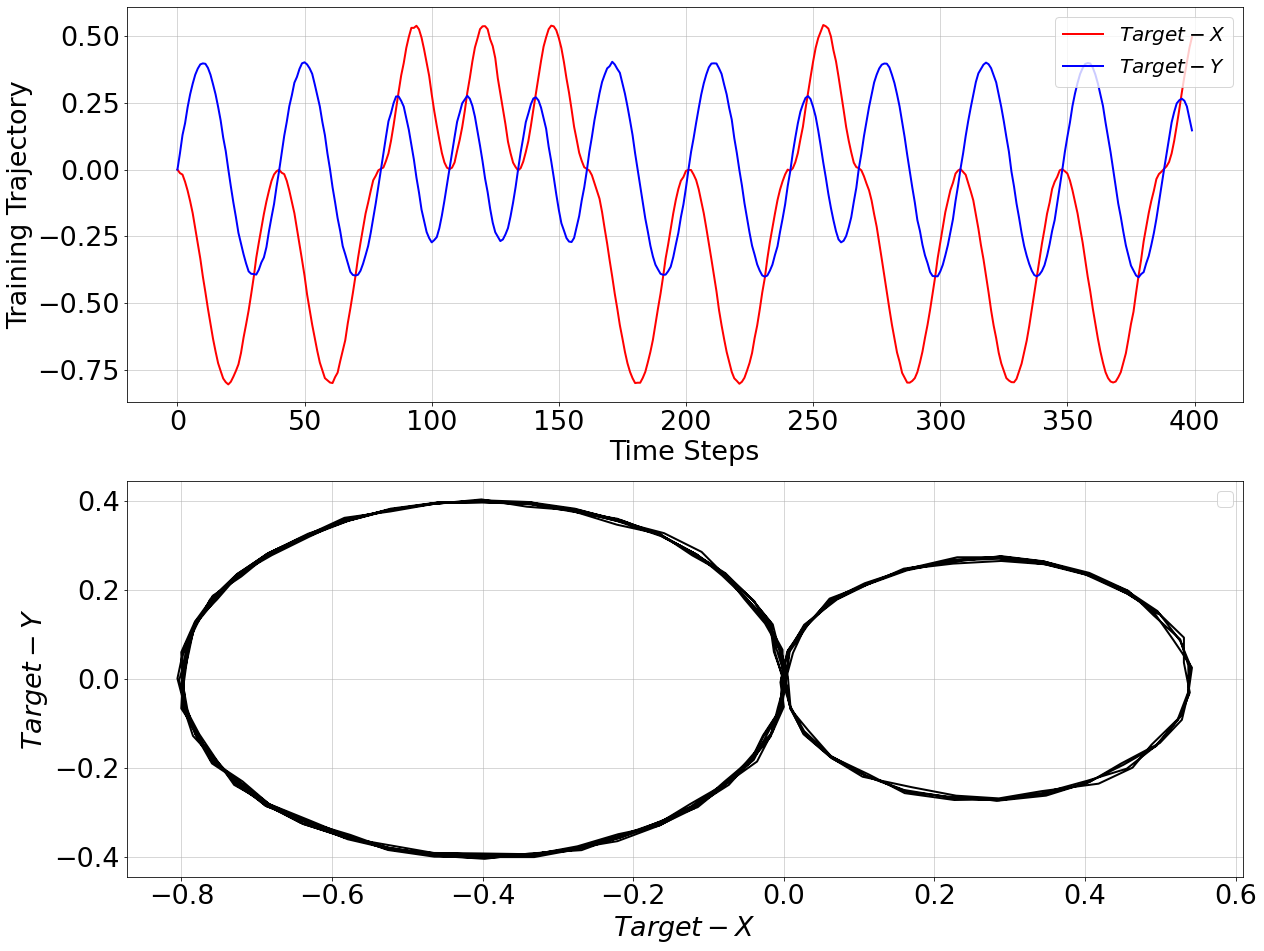}
\caption{Training trajectory over 400 time steps (top plot) and its representation in $X-Y$ space (bottom plot). $Target - X$ and $Target - Y$ correspond to the first and second dimensions of the training trajectory, respectively. 
\label{fig:training_traj}}
\end{figure}

First, we trained a PV-RNN with 2-dimensional sensory sequence data. 
The training data comprised 80 sequences, each containing 2160 time steps. For preparing those trajectories, we designed 2 different 2-dimensional cyclic patterns, one with periodicity of 40 time steps and the other with periodicity of 27 time steps. Each trajectory was made of probabilistic switching among these 2 cyclic patterns wherein after one cycle of a particular pattern the same pattern repeats with a probability of 60\% and the pattern transits to the other pattern with a probability of 40\% equally. Noise has been added to individual points at randomly spaced intervals. The intervals between noise points are determined by drawing from a normal distribution (mean of 1, standard deviation of 10), providing a variable time step size. At each noise interval, Gaussian noise (mean of 0, standard deviation of 0.001) is added to the current data point, slightly perturbing its coordinates to simulate natural fluctuations without disrupting the cyclic structure.
A part of the training trajectory is shown in Figure \ref{fig:training_traj}. 

The network parameters used for training PV-RNN are listed in Table \ref{tab:parameters}. $\#\mathbf{d}$, $\#\mathbf{z}$, $\tau$, $\mathbf{w}^{tr}$ indicates the number of $\mathbf{d}$ neurons, number of $\mathbf{z}$ neurons, time constant, and meta-prior during the training phase, respectively.
%
% TABLE: network parameters
%
\begin{table}[H]
\caption{PV-RNN training parameters.}
\label{tab:parameters}
\scalebox{1}[1]{
\begin{tabularx}{\textwidth}{
>{\hsize=1\hsize}X % First column: smaller
>{\hsize=1\hsize}X   % Second column: standard
>{\hsize=1\hsize}X   % Third column: standard
>{\hsize=1\hsize}X   % Fourth column: standard
>{\hsize=1\hsize}X % Fifth column: larger
>{\hsize=1\hsize}X % Sixth column: larger
}
\toprule
     & $\#\mathbf{d}$ & $\#\mathbf{z}$ & \boldmath{$\tau$} & $\mathbf{w}^{tr}$ \\ %& $\mathbf{w}^{ln}$ \\
\midrule
\textbf{Layer 1} 
& 30 
& 2 
& $1$  
& $0.001$ \\
%& $100$ \\
\textbf{Layer 2} 
& 15 
& 1 
& $5$ 
& $0.01$ \\
%& Layer 1 $\mathbf{w}^{tr} \times 10$ \\
%& Layer 1 $\mathbf{w}^{ln} \times 10$ \\
\bottomrule
\end{tabularx}}
\end{table}

The PV-RNN was trained over $130,000$ epochs minimizing free energy in Equation \ref{eq:loss} using the Adam optimizer \cite{kingma2014adam} and back-propagation through time (BPTT) \cite{rumelhart1985learning} with learning rate $0.001$ to optimize all network parameters of $\theta$ and $\phi$ of the generative and inference model, and the adaptive variable $\mathbf{A}$ corresponding to each training trajectory.

\begin{figure}[H]
\centering
\includegraphics[width=0.9\textwidth]{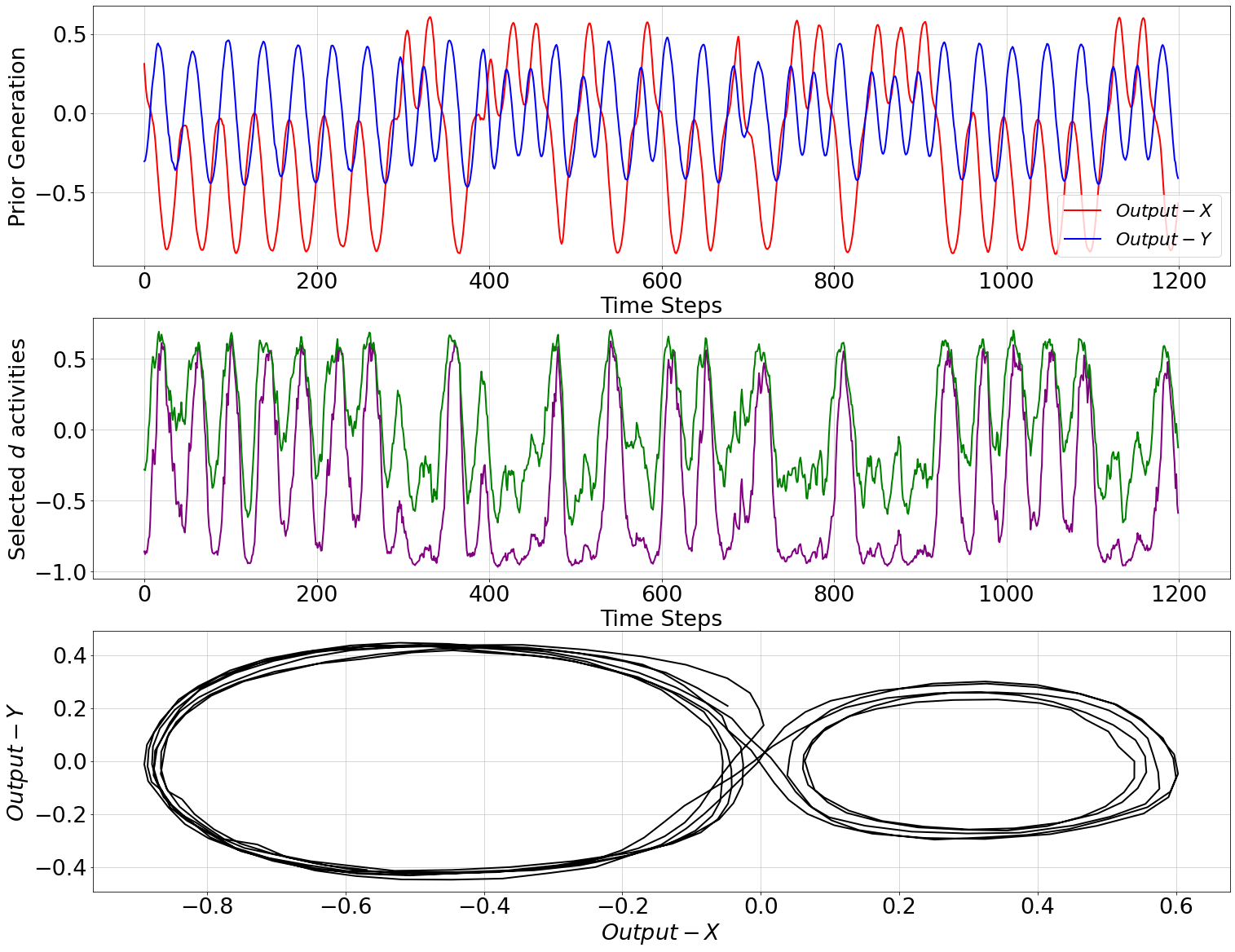}
\caption{Prior generation over 1200 time steps under trained model with meta-prior $\mathbf{w}^{tr}$ from Table \ref{tab:parameters} (top plot), selected activities of the $\mathbf{d}$ neurons in the bottom layer of the PV-RNN (middle plot), and a representation in $X-Y$ space (bottom plot). $Output - X$ and $Output - Y$ correspond to the first and second dimensions of the prior generation output trajectory, respectively.
\label{fig:prior_generation}}
\end{figure}

The trained network was evaluated on the basis of how well probabilistic transitions in the training data were reflected in the PV-RNN generative process, the so-called prior generation of the PV-RNN, which is conducted without performing the inference of the latent variables with sensory observation. 
In prior generation, the prior distribution $\mathbf{z}^p_{1}$ was initialized with a unit Gaussian (Equation \ref{eq:p}) and then latent states were recursively computed to generate network output sequences. 
Figure \ref{fig:prior_generation} shows an example of the prior generation outputs over 1200 time steps. We can see that the patterns shift from one to another, where the two patterns used for training appear randomly. In addition, using a categorizer to discriminate between the two patterns, different prior generation outputs over 50,000 time steps have shown a probability of 38\%-43\% of switching to a different pattern and a probability of 57\%-62\% of staying in the same pattern, which are close to the training dataset.

\subsection{Testing of perception task}\label{sec:results:testing}
The trained PV-RNN was tested by performing the perception task.
In the test, the inference process was performed within the inference window, while one of the trained patterns was used as the target sensory sequence for the inference of the latent variables. The length of the inference window was set to 400 time steps. The adaptation of meta-prior, $\mathbf{w}$, during inference with the monitoring of the average prediction error over 300 time steps                                                               
\footnote{For the implementation strategy described in Algorithm \ref{alg:adaptive_switching}, the length of the inference window and time window for computing the average reconstruction error can be considered design decisions and do not need to be the same length. This choice may depend on how the high and low thresholds are defined, which impact the probability of transition from FS to MW (and vice-versa) and, thus, the expected probabilistic behavior of the system. For instance, an average reconstruction error computed over a small time window may not reach or may be too far from a desired threshold. In this scenario, the probability of staying in the current state (FS or MW) would remain large over the entire simulation time.}                                          was carried out using the parameters listed in Table \ref{tab:param_test}. 
% where $\mathbf{w}^{Low}$ and $\mathbf{w}^{High}$ correspond to the meta-prior under Focus State (FS) and Mind-Wandering (MW) episode, respectively.
%
% TABLE: network parameters
%
\begin{table}[H]
\caption{PV-RNN testing parameters.}
\label{tab:param_test}
\scalebox{1}[1]{
\begin{tabularx}{\textwidth}{
>{\hsize=1\hsize}X % First column: smaller
>{\hsize=1\hsize}X   % Second column: standard
>{\hsize=1\hsize}X|  % Third column: standard
>{\hsize=1\hsize}X   % Fourth column: standard
>{\hsize=1\hsize}X % Fifth column: larger
>{\hsize=1\hsize}X % Sixth column: larger
}
\toprule
      & $\mathbf{w}^{L}$ & $\mathbf{w}^{H}$ & $\mathbf{Temp}$ & $\mathbf{Thr_L}$ & $\mathbf{Thr_H}$ \\
\midrule
\textbf{Layer 1} 
%& 30 
%& 2 
%& $1$  
& $0.001$ 
& $100$ 
& 0.01
& 0.1
& 0.5 \\
\textbf{Layer 2} 
%& 15 
%& 1 
%& $5$ 
& $0.01$ 
& $1000$ \\
\bottomrule
\end{tabularx}}
\end{table}

\begin{figure}[H]
\centering
\includegraphics[width=\textwidth]{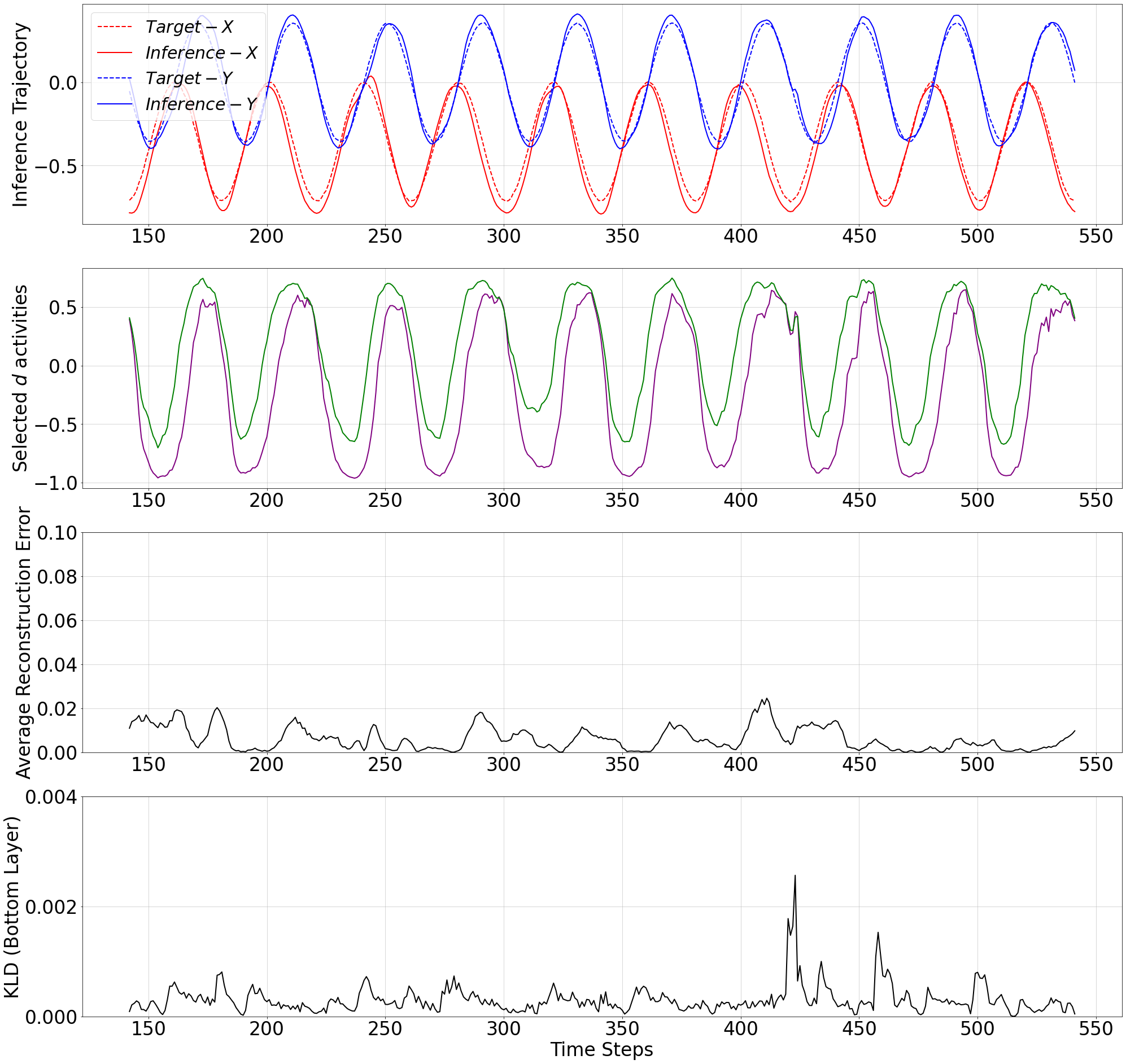}
\caption{From top to bottom: inference output trajectory with meta-prior $\mathbf{w}^{L}$ from Table \ref{tab:param_test}, selected activities of the $\mathbf{d}$ neurons in the bottom layer of the PV-RNN, average reconstructions error over the inference window at time step 542, and KL divergence at the PV-RNN bottom layer. $Inference - X$ and $Inference - Y$ correspond to the first and second dimensions of the inference output trajectory, respectively.
\label{fig:test_plots_FS}}
\end{figure}

\begin{figure}[H]
\centering
\includegraphics[width=\textwidth]{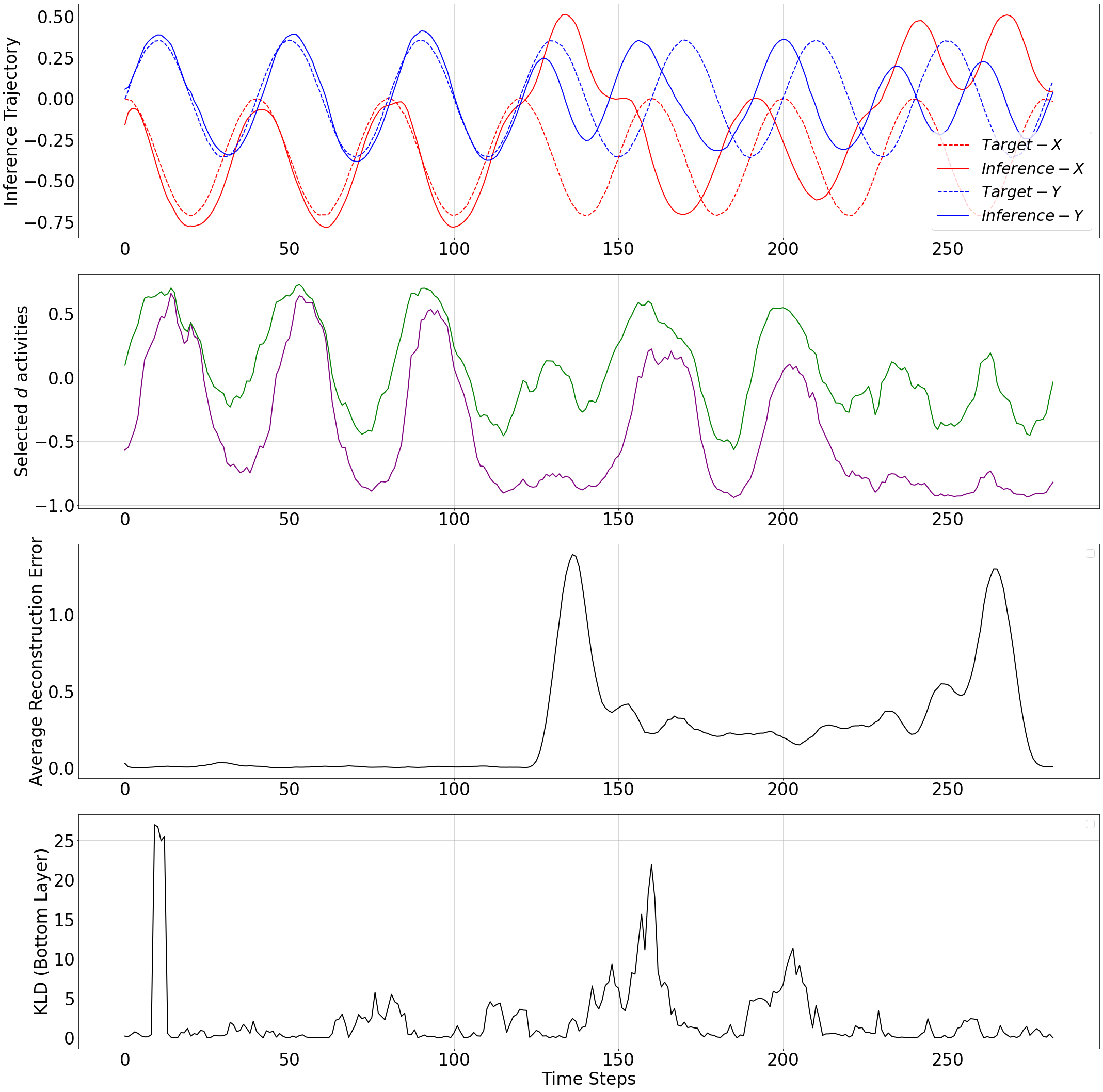}
\caption{From top to bottom: inference output trajectory with meta-prior $\mathbf{w}^{H}$ from Table \ref{tab:param_test}, selected activities of the $\mathbf{d}$ neurons in the bottom layer of the PV-RNN, average reconstructions error over the inference window at time step 283, and KL divergence at the PV-RNN bottom layer. $Inference - X$ and $Inference - Y$ correspond to the first and second dimensions of the inference output trajectory, respectively.
\label{fig:test_plots_MW}}
\end{figure}

The mechanistic behavior when $\mathbf{w}$ adapted to low and high values are shown in Figures \ref{fig:test_plots_FS} and \ref{fig:test_plots_MW}, respectively.
The plots show the output trajectory, the target sensory sequence, the average prediction error, and KL divergence at the PV-RNN bottom layer for each case.
It can be seen in Figure \ref{fig:test_plots_FS} that when $\mathbf{w}$ adapted to $\mathbf{w}^{L}$, a pattern used for the target sensory sequence is generated well during inference while the average prediction error remains low (below 0.03) over the entire inference window. 
This indicates that adaptation of $\mathbf{w}$ to $\mathbf{w}^{L}$ enabled the output to accurately reconstruct the target sensory sequence. This period is analogous to a situation of FS.

On the other hand, Figure \ref{fig:test_plots_MW} demonstrates a period when $\mathbf{w}$ adapted to $\mathbf{w}^{H}$. 
In this period, the inference trajectory is generated similarly to the prior generation shown in (Figure \ref{fig:prior_generation}). 
In particular, we can observe in Figure \ref{fig:test_plots_MW} that after a few cycles of one pattern, the inference trajectory generates the other pattern, returns to the previous pattern (which is out of phase from the target sensory sequence due to the different periodicity between the two cyclic patterns), and then switches again to the other pattern. 
As a result, the average reconstruction error increases once the inference trajectory starts to deviate from the target sensory sequence.
This observation is analogous to a situation of MW.

Selected $\mathbf{d}$ activities of the PV-RNN bottom layer during prior generation after training, as well as during inference with adaptation of $\mathbf{w}$ to low and high values, are shown in Figures \ref{fig:prior_generation}-\ref{fig:test_plots_MW}. 
In both cases of the inference, the correspondence between $\mathbf{d}$ activity patterns and the output patterns is analogous to that observed during prior generation after training. Specifically, the $\mathbf{d}$ activities follow a single pattern when $\mathbf{w}$ adapted to the low value, while the $\mathbf{d}$ activities alternate between two patterns, closely reflecting the dynamics of the $\mathbf{d}$ activities seen in the prior generation when $\mathbf{w}$ adapted to the high value.

\begin{figure}[H]
\centering
\includegraphics[width=\textwidth]{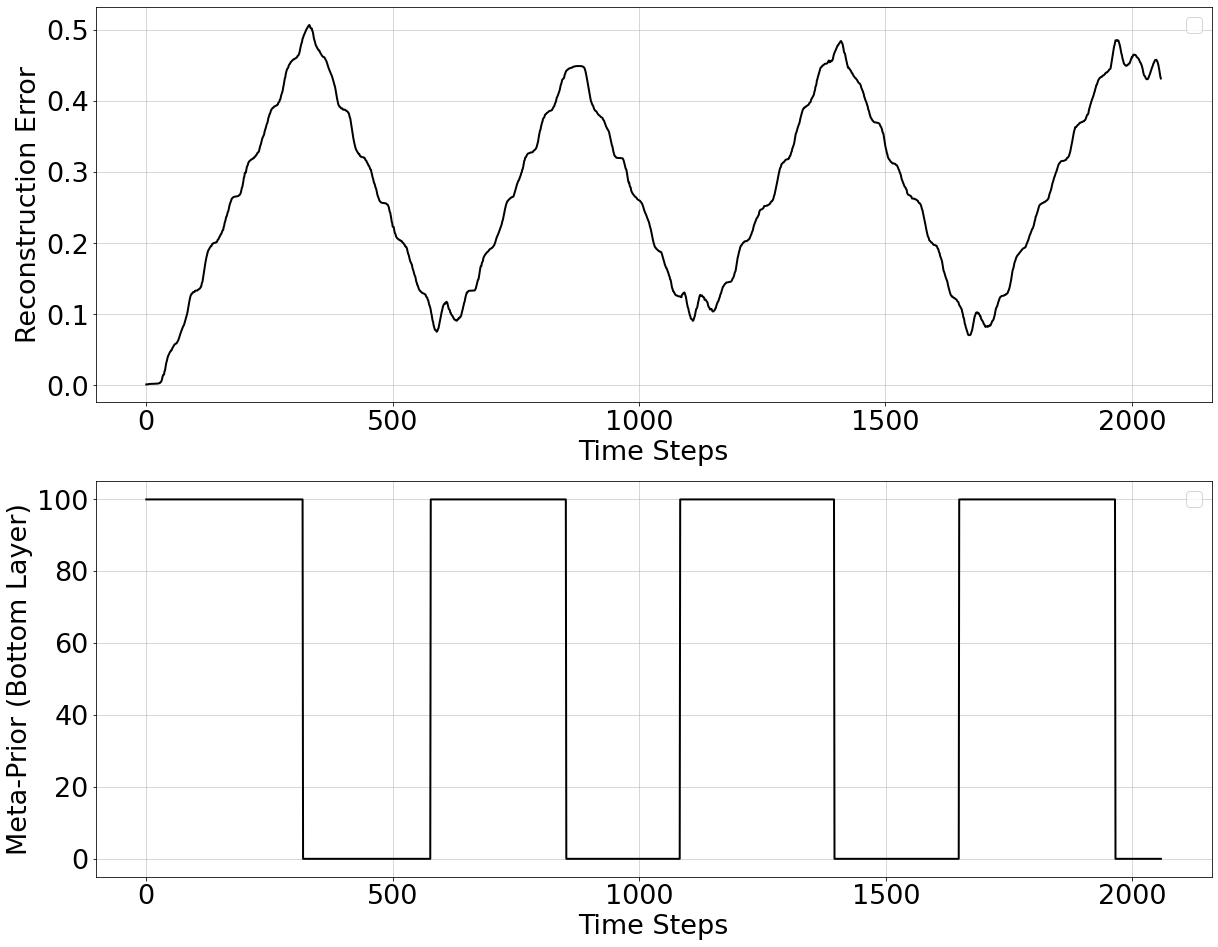}
\caption{Reconstruction error over inference window computed at each time step (top plot) and adaptive meta-prior value ($\mathbf{w}$) of the PV-RNN bottom layer over time (bottom plot). 
%Meta-prior of 0.001 and 100 represent a focus state (FS) and mind-wandering (MW) phase, respectively.
\label{fig:recc_w_plots}}
\end{figure}

% The overall behavior described above under the proposed autonomous shifts between two distinct periods can be visualized in Figure \ref{fig:recc_w_plots}. 
Figure \ref{fig:recc_w_plots} shows the overall behavior of autonomous shifts between two distinct periods obtained in the experiments.
The plots show the average reconstruction error during inference and meta-prior values of the PV-RNN bottom layer over time.
It can be observed that when PV-RNN is under $\mathbf{w}^{H}$, the average prediction error increases as close to the high threshold value, which makes the probability of switching from $\mathbf{w}^{H}$ to $\mathbf{w}^{L}$ larger according to equation \ref{eq:adaptW_MW}. 
Then, $\mathbf{w}$ is switched to the low value ($\mathbf{w}^{L}$). After this shift, the average prediction error continues to decline until it becomes close to the low threshold value, which increases the probability of switching from $\mathbf{w}^{L}$ to $\mathbf{w}^{H}$. 
$\mathbf{w}$ is then switched back to the high value. 
The former case corresponds to the shift from MW to FS and the latter case corresponds to the shift from FS to MW.

\begin{figure}[H]
\centering
\includegraphics[width=\textwidth]{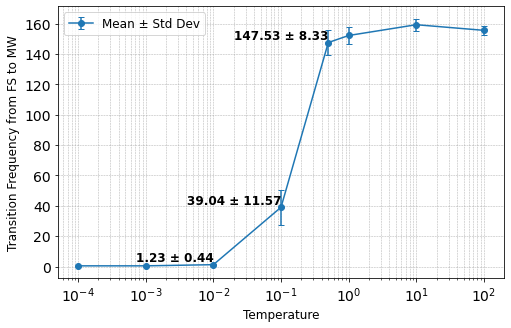}
\caption{Transition frequency from FS to MW per 1000 time steps under different temperature values. The mean and standard deviation are displayed for three intermediate cases when temperature is 0.01, 0.10, and 0.50.
\label{fig:freq_temp_plot}}
\end{figure}

Finally, we investigated the effect of changing temperature values on the characteristics of the shifts between FS and MW. 
For this purpose, we counted the number of transitions occurred from FS to MW during 1000 steps in the perception test.
The results are shown in Figure \ref{fig:freq_temp_plot}. 
It can be seen that the transition frequency from FS to MW increases when the temperature increases. In particular, for larger temperature values, the transitions from FS to MW become more frequent (i.e., the system becomes more random) since the probability of switching from FS to MW becomes closer to 50\% due to the argument inside the sigmoid function being closer to zero in equation \ref{eq:adaptW_F}. In contrast, when the temperature is smaller, the transitions from FS to MW become less frequent (i.e., the system becomes more deterministic), which primarily happen when the average reconstruction error reaches the low threshold.                           
%\footnote{Similar to the analysis of the impact of changing temperature values on the characteristics of the shifts between FS and MW, one could also make an analysis of the impact of the high and low threshold values given a fixed temperature. In particular, it is expected that the transition frequency from FS to MW increases when the difference between the high and low threshold values is decreased.}               
For the case study in Figure \ref{fig:recc_w_plots}, 0.01 was chosen to be the temperature with a mean of 1.23 transitions from FS to MW per 1000 time steps.

%===================================================
%===================================================
%\input{sections/4_Discussion}
\section{Discussion}\label{sec:4}
This study explored the neural mechanisms underlying autonomous shifts between the focused state (FS) and mind-wandering (MW) through simulation experiments using a newly proposed model based on the free energy principle. 
The proposed model, an extension of PV-RNN, introduces an adaptation mechanism for a meta-level parameter, the meta-prior $\mathbf{w}$, which is modulated based on the average reconstruction error over a fixed-size past window. 
Specifically, $\mathbf{w}$ probabilistically switches to a high value when the average reconstruction error decreases close to a minimal threshold and to a low value when the average reconstruction error increases near a maximal threshold.

In the simulation experiments, the PV-RNN was first trained to generate probabilistic transitions between two distinct cyclic patterns. 
% In the perception task phase, the inference process was performed while inferring latent variables within the inference window by observing one of the trained cyclic patterns wherein Autonomous shifts between FS and MW were observed, accompanied by dynamic changes in $\mathbf{w}$. 
In the perception task phase, latent variables within the inference window were inferred to minimize the reconstruction error for given target sensory sequence while adapting $\mathbf{w}$.
One of the trained cyclic patterns was used as the target.
% The meta-prior values during inference described in Table \ref{tab:param_test} were chosen based on how well the inference process of the PV-RNN can adapt and reconstruct the sensory target sequence ($\mathbf{w}$ = $\mathbf{w}^{L}$), resulting in small average reconstruction error, or computed similarly to the prior generation outputs after training ($\mathbf{w}$ = $\mathbf{w}^{H}$), resulting in small KL divergence values but larger average reconstruction error.

When $\mathbf{w}$ shifted to a low value, stronger bottom-up sensory perception dominated, regenerating the observed sensory sequence in the outputs with minimal reconstruction error while allowing larger Kullback-Leibler divergence between the prior and the approximate posterior. 
This leads to a focused state. 
% Conversely, when $\mathbf{w}$ shifted to a high value, the divergence between the inference trajectory and the prior (indicated by an increase in the Kullback-Leibler divergence) allowed stronger top-down processing, resulting in a state resembling mind-wandering (Figure \ref{fig:test_plots_MW}).
Conversely, when $\mathbf{w}$ shifted to a high value, the approximated posterior is attracted toward the prior by stronger mean of minimizing the Kullback-Leibler divergence between the prior and the approximated posterior.
This allowed stronger top-down processing while less attending to sensation, generating relatively large reconstruction error in the inference window.
This results in a state resembling mind-wandering.

One limitation of the current study is that the proposed model does not account for the phenomenon of becoming consciously aware of MW, which enables redirection of attention back to FS. 
\cite{sandved2021towards} hypothesize that inferring a "true meta-state" by asking, "How aware am I of where my attention is?" could trigger self-awareness of MW. 
While the dynamically changing meta-prior in the current model modulates the balance between top-down and bottom-up information flow, leading to shifts between FS and MW, it may correspond to the meta-state proposed in \cite{sandved2021towards}. 
However, the current model lacks a mechanism for explicitly inferring such a meta-state, making it unable to account for self-awareness of it. 
Future studies should address this limitation by extending the model to include an inference mechanism for a meta-state.

How do the current results relate to either the two-stage model \cite{voss2018new} or the multiple sub-event model \cite{zukosky2021spontaneous} described previously? 
The two-stage model suggests that the probability of remaining in FS decreases over time during an FS-MW episode, which concludes with conscious awareness of MW. 
In contrast, the multiple sub-event model posits a lesser decrease in this probability, speculating that multiple unconscious shifts between FS and MW occur before MW is consciously noticed. 
Since the current model does not account for self-awareness of MW, as discussed earlier, it is challenging to directly align its results with either of these models.

Finally, numerous studies have indicated that MW during the resting state is intricately linked to the functional organization and dynamics of brain networks, particularly the default network (DN), central executive network (CEN), and salience network (SN) \cite{mason2007wandering,godwin2017functional,denkova2019dynamic}.The current study does not model interactions between such distinct networks. 
Extending the model to incorporate dynamic interactions among these networks would provide a tighter connection to established neuroscientific findings on resting-state phenomena and offer deeper insights into mind-wandering.

%===================================================
%===================================================

%\section*{Acknowledgments}
%This study was funded by the Japan Society for the Promotion of Science (JSPS) KAKENHI, Transformative Research Area (A): unified theory of prediction and action [24H02175].

%Bibliography
\bibliographystyle{unsrt}  
%\bibliography{manuscript} 
\bibliography{library }
%=====================================
%=====================================

\end{document}